\documentclass[11pt,a4paper]{article}
\pdfoutput=1
\usepackage{amsfonts}
\usepackage{tikz}
\usepackage{tabularx}
\newcolumntype{C}{>{\centering}X}

\def\tp{\otimes}
\def\ds{\oplus}
\def\ket[#1]{\left| #1 \right>}
\def\bra[#1]{\left< #1 \right|}
\def\L{\mathcal{L}}
\def\C{\mathbb{C}}

\title{From spin to anyon notation: The XXZ Heisenberg model as a $D_{3}$ (or $su(2)_{4}$) anyon chain}
\author{Peter E. Finch \\
  \\
  Institut f\"ur Theoretische Physik, \\
  Leibniz Universit\"at Hannover, \\
  Appelstra\ss e 2, 30167 Hannover, Germany}
\date{}

\begin{document}
\maketitle

\begin{abstract}
We discuss a relationship between certain one-dimensional quantum spin chains and anyon chains. In particular we show how the XXZ Heisenberg chain is realised as a $D_{3}$ (alternately $su(2)_{4}$) anyon model. We find the difference between the models lie primarily in the choice of boundary conditions.
\end{abstract}

\section{Introduction}
Formulations of many-body systems considered in quantum information theory differ from what is traditionally used in condensed matter physics. It is important that the different approaches are compared and that communication between the fields occurs. Here we aim to shed light on and discuss the relationship between the recently proposed anyon chains from topological quantum information \cite{FTLTKWF2007} and their quantum spin chain cousins from condensed matter physics. Specifically we will provide conditions which when met imply a complete equivalence between the two different formalisms.

The one-dimensional anyon chains were constructed analogously to spin chains to provide a stepping stone to the understanding of higher dimensional anyonic models. These models were successful at demonstrating the presence of topological symmetries \cite{AGLTT2011,FTLTKWF2007}. The anyons used to construct these models typically exhibit non-standard braiding statistics and are not required to have integer quantum dimension\footnote{Roughly speaking, the quantum dimension is the dimension of the internal Hilbert space of the particle and determines the probability that fusion leads to annihilation or the creation of other anyons \cite{Preskill2004}.}.
The global Hilbert spaces used for the anyon chains rely on fusion paths \cite{MooSei1989,Preskill2004} and often have no tensor product structure. One fruitful method of constructing anyonic theories utilises quasi-triangular Hopf algebras \cite{WilBai1998}. These anyon chains can also be constructed as the Hamiltonian limits of interaction-round-a-face (IRF) or restricted solid-on-solid (RSOS) models, albeit not restricted to the ADE classification.

On the other hand there are quantum spin chains which have a well established place in modern condensed matter physics, providing insight into critical behaviour of correlated physical systems and describing quasi one-dimensional materials \cite{EFGKVBook2005}. As we are discussing their connection to anyon chains we will consider spin chains that have the underlying symmetry of a quasi-triangular Hopf algebra (e.g. a quantum group) \cite{ChaPreBook1994,MajidBook1995}. Consequently, it is natural to discuss models constructible from the Quantum Inverse Scattering Method (QISM) \cite{KBIBook1993,STF1979} and its variants, although all results will have no dependency on integrability. 

An equivalence between spin and anyon chains occurs when the underlying symmetry of each is that of the same quasi-triangular Hopf algebra. This equivalence has previously appeared as a face-vertex correspondence for integrable two-dimensional lattice models \cite{Pasquier1988,Roche1990}. The correspondence for one-dimensional quantum models will be illustrated by presenting the nearest-neighbour XXZ Heisenberg chain viewed as a $D_{3}$ anyon model.
We use this model as it is a well-known and simple model for which we can calculate operators and energy spectra explicitly. The generalisation to other models with Hopf algebra symmetries is straight-forward.

While the local XXZ Hamiltonian has the complete $D_{3}$ symmetry, the symmetry of the global Hamiltonian depends upon the boundary conditions imposed. Thus the correspondence depends upon the boundary conditions; an aspect of these models not previously discussed.\footnote{While this correspondence has not been discussed in terms one-dimensional quantum chains, similar observations relating to the partition function of certain two-dimensional lattice models have been made e.g. \cite{VegGia1989}.}
We consider open boundaries with free ends \cite{ABBBQ1987,Sklyanin1988}, periodic boundaries of both spin \cite{CloGau1966,Orbach1958,YanYan1966a} and anyon type \cite{FTLTKWF2007,TTWL2008}, and braided boundaries \cite{GPPR1994,KarZap1994,LinFoe1997}. Of these only the open and braided boundary conditions always have an equivalent description in the spin and anyon pictures.

It is also possible to present the XXZ model using other underlying symmetries, e.g. $su(2)_{4}$, $D_{5}$ or $U_{q}(su(2))$, however, using $D_{3}$ has certain advantages. There are no superfluous anyons, like the anyons in half-integer subsector of $su(2)_{4}$ or an additional anyon of quantum dimension two in $D_{5}$.  The anisotropy parameter is not dependent upon the algebra like $U_{q}(su(2))$ where $\frac{J_{z}}{J} = \cosh(\ln(q))$ ($J$ and $J_{z}$ are the coupling constants of the model).  We also note that the XXZ Heisenberg chain has appeared in other papers in anyonic form, specifically as the spin-1 $su(2)_{4}$ chain \cite{GATHLTW2012,VMP2011}, although not discussed as such.

\section{Background}
Here we present the background information for the XXZ Heisenberg model, the $D_{3}$ algebra and the spin and anyon bases for the models. We also discuss when the operators in each of the bases are said to have the symmetry of $D_{3}$.\\

\noindent
\textbf{\underline{The algebra}} \\
\noindent
$D_{3}$ is the group of symmetries on a triangle consisting of a rotation, $\sigma$, and flip, $\tau$. 
The group has the presentation,
$$ D_{3} = \{\sigma,\tau | \sigma^{3} = \tau^{2} = \sigma \tau \sigma \tau = 1 \}. $$
Its group algebra is the linear combination of its elements over the complex numbers. It is also possible to embed this algebra into a $k$-fold space by use of the general coproduct,
$$ \Delta^{(k)}(g) = \overbrace{g \tp ... \tp g}^{k-\mbox{times}}, \hspace{2cm} g \in D_{3}, $$
extended linearly to the algebra. This algebra is known to form a quasi-triangular Hopf algebra \cite{ChaPreBook1994,MajidBook1995}. As it is cocomutative the universal $R$-matrix is just the identity operator. The representation theory of this algebra is also known, it has three irreps (irreducible representations), two are one-dimensional,
$$ \pi_{\pm}(\sigma) = 1, \hspace{0.5cm} \pi_{\pm}(\tau) = \pm 1, $$
and one is two-dimensional,
$$ \pi_{2}(\sigma) = \left(\begin{array}{cc} e^{\frac{2i\pi}{3}} & 0 \\ 0 & e^{-\frac{2i\pi}{3}} \end{array}\right), \hspace{0.5cm} \pi_{2}(\tau) = \left(\begin{array}{cc} 0 & 1 \\ 1 & 0 \end{array}\right). $$
For each irrep $\pi_{a}$ we will associate a space (module) $V_{a}$. The fusion rules are as follows:
$$ V_{-} \tp V_{-} \cong V_{+}, \hspace{1cm} V_{-} \tp V_{2} \cong V_{2}, \hspace{1cm} \mbox{and} \hspace{1cm} V_{2} \tp V_{2} \cong V_{2} \ds V_{+} \ds V_{-}. $$
The space $V_{+}$ is the vacuum or trivial space and fusion with it is trivial, i.e. $V_{+}\tp V_{a}\cong V_{a}$. As fusion is associative i.e. $(V_{a} \tp V_{b}) \tp V_{c} \equiv V_{a} \tp (V_{b} \tp V_{c})$, there exists $F$-moves (generalised 6-j symbols) which relate the two different ways to decompose the 3-fold tensor product space \cite{MooSei1989,TTWL2008}. The interpretation of these operators can be understood via the diagrammatic interpretation presented later. The $F$-moves can be explicitly constructed from the representations above and are found in Appendix B. 

To construct an anyonic model we associate with each irreducible $D_{3}$ module an anyon \cite{WilBai1998}. The fusion of anyons is governed by the fusion of the respective modules. Similarly the other properties of the anyons are inherited from the representation theory of the algebra.\footnote{As the quantum dimension of the anyon equals the dimension of the associated irrep all the anyons will have integer quantum dimension. This will also be true for other anyonic theories defined via matrix representations of Hopf algebras.} \\

\noindent
\textbf{\underline{The Bases}} \\
\noindent
The Hilbert space for the spin formalism consists of the tensor product of $\mathcal{L}$ sites containing spin-$\frac{1}{2}$ particles, alternately qubits or other 2-state systems, coupled to an auxiliary $\C^{4}$ space, specifically
$$ [V_{2} \ds V_{+} \ds V_{-}] \tp V_{2}^{\tp \mathcal{L}}. $$ 
This has a natural basis with $2^{\mathcal{L}+2}$ vectors, which we shall refer to as the spin basis. We also note that if we project onto the $V_{+}$ component of the auxiliary space then we are just left with $\mathcal{L}$ spin-$\frac{1}{2}$ sites.\footnote{For more general models the global Hilbert space will be $\L$ copies of a certain irrep, corresponding to an anyon, coupled to an auxiliary space which is the direct sum of all irreps, i.e. a reservoir of all anyons.}

To form the Hilbert space for the anyon formalism we need to consider fusion paths \cite{MooSei1989,Preskill2004} and as such we will refer to its basis as the \textit{fusion path basis}. As this Hilbert space will be equivalent to the spin formalism we again have an auxiliary space of $V_{2} \ds V_{+} \ds V_{-}$ followed by $\L$ copies of $V_{2}$. To form a fusion path, working from left to right, we first fuse an irreducible subspace in the auxiliary space with $V_{2}$, choosing which irreducible space we project onto. This is then fused to another $V_{2}$, so on and so forth. We record the irreducible subspace of the auxiliary space and the subsequent irreducible subspaces which appear after fusion,
\vspace{0.2cm}
$$ \begin{array}{l}
	\begin{tikzpicture}[scale=1.0]
	\put (30,0){$2$}	\put (60,0){$2$}	\put (90,0){$2$}	\put (150,0){$2$}	\put (180,0){$2$}
	\put (0,-20){$a_{0}$}	\put (40,-27){$a_{1}$}	\put (70,-27){$a_{2}$}	\put (100,-27){$a_{3}$}	\put (160,-27){$a_{\mathcal{L}-1}$}	\put (205,-20){$a_{\mathcal{L}}$}
	\draw (0.5,-0.6) -- (4.0,-0.6);	\draw (4.2,-0.6) -- (4.4,-0.6);	\draw (4.6,-0.6) -- (4.8,-0.6);	\draw (5.0,-0.6) -- (7.0,-0.6);
	\draw (1.15,-0.1) -- (1.15,-0.6);	\draw (2.20,-0.1) -- (2.20,-0.6);	\draw (3.25,-0.1) -- (3.25,-0.6);	\draw (5.35,-0.1) -- (5.35,-0.6);	\draw (6.40,-0.1) -- (6.40,-0.6);
\end{tikzpicture} \vspace{0.6cm} \\
	\hspace{0.1cm} \equiv \hspace{0.1cm} ((\cdot((V_{a_{0}} \tp V_{2})_{a_{1}} \tp V_{2})_{a_{2}} \cdot\cdot\cdot )_{a_{\mathcal{L}-1}} \tp V_{2})_{a_{\mathcal{L}}} \vspace{0.4cm} \\
	\hspace{0.1cm} \equiv \hspace{0.1cm} \ket[a_{0}a_{1}a_{2}...a_{\mathcal{L}-1}a_{\mathcal{L}}].
\end{array} $$
Here the use of the subscript of the bracket, $(\cdot\cdot)_{a}$, denotes the subspace of the space inside the bracket isomorphic to $V_{a}$. These sequences of labels form the basis vectors of the anyon Hilbert space. Using this formulation we observe these fusion path basis vectors correspond to a subspace in the spin basis, whose dimensionality is equal to that of dimension of the out-going anyon.\footnote{While an individual fusion path basis vector will correspond to a subspace in the spin basis, generic vectors in the anyon Hilbert space will have no such correspondence.} We remark that each label $a_{i}$, as it is produced by fusion, is limited by the preceding label. We find that neighbouring pairs must belong to the following set,
$$ a_{i}a_{i+1} \in \{ +2, -2, 2+, 2-, 22 \}. $$
We now have an appropriate fusion path basis. 

Diagrammatically we have fused left to right, however we can rearrange fusion adopting the additional convention of also fusing top to bottom. The reordering of fusion is done by the aforementioned $F$-moves. In terms of fusion diagrams we have,
\vspace{0.3cm}
\begin{eqnarray*}
\begin{tikzpicture}[scale=1.0]
	\put (26,5){$b$}	\put (56,5){$c$}
	\put (0,-20){$a$}	\put (40,-27){$d$}	\put (83,-20){$e$}
	\draw (0.35,-0.6) -- (2.85,-0.6);
	\draw (1.00,0.1) -- (1.00,-0.6);	\draw (2.05,0.1) -- (2.05,-0.6);
\end{tikzpicture} \hspace{0.3cm} 
& = & 
\sum_{d'} (F^{abc}_{e})^{d}_{d'} \hspace{0.5cm}
\begin{tikzpicture}[scale=1.0]
	\put (26,5){$b$}	\put (56,5){$c$}
	\put (8,-20){$a$}	\put (47,-15){$d'$}	\put (76,-20){$e$}
	\draw (0.60,-0.6) -- (2.60,-0.6);
	\draw (1.00,0.1) -- (1.525,-0.25);	\draw (2.05,0.1) -- (1.525,-0.25); 	\draw (1.525,-0.25) -- (1.525,-0.6);
\end{tikzpicture}
\end{eqnarray*}
On the left the anyons $a$ and $b$ are fused with the result fusing to the anyon $c$, while on the right the anyons $b$ and $c$ are fused with the result fusing to $a$. The $F$-moves must satisfy a pentagon equation, although in the $D_{3}$ case this is automatically satisfied as $D_{3}$ forms a Hopf algebra.

We also want to determine the dimensionality of the anyon Hilbert space. We define the number $N_{\mathcal{L}}^{(ab)}$ to be the number of basis vectors with $a_{0}=a$ and $a_{\mathcal{L}}=b$. The numbers are determined by the relations
$$ N_{\mathcal{L}}^{(a+)}  = N_{\mathcal{L}}^{(a-)} = N_{\mathcal{L}-1}^{(a2)}, \hspace{0.7cm} N_{\mathcal{L}}^{(ab)} = N_{\mathcal{L}}^{(ba)} \hspace{0.7cm} \mbox{and} \hspace{0.7cm} N_{\mathcal{L}}^{(a2)} = N_{\mathcal{L}-1}^{(a2)} + 2N_{\mathcal{L}-1}^{(a+)}. $$
We can recognise that these numbers are those appearing the in Jacobsthal sequence (A001045 \cite{OEIS}) which have the form
$$ N^{(22)}_{\mathcal{L}} = \frac{1}{3}\left( 2^{\mathcal{L}+1}+(-1)^{\mathcal{L}} \right). $$
Using this we can determine the dimension of the anyon Hilbert space,
$$ \mbox{Anyon dimension} = \sum_{a,b} N_{\mathcal{L}}^{(ab)} = N_{\mathcal{L}+2}^{(22)} = \frac{1}{3}\left( 2^{\mathcal{L}+3}+(-1)^{\mathcal{L}} \right) .$$
We can also determine that we indeed have the correct dimension of the spin Hilbert space
$$ \mbox{Spin dimension} = \sum_{a,b} N_{\mathcal{L}}^{(ab)} \times \mbox{dim}(V_{b})= 2^{\mathcal{L}+2}. $$ \\

\noindent
\textbf{\underline{Equivalent Operators}} \\
\noindent
The global Hilbert spaces were both constructed using $D_{3}$ models and therefore they should be $D_{3}$ modules themselves.
It follows that for an operator to be expressible in both the spin and anyon bases it must have the underlying symmetry of the $D_{3}$ algebra. In the spin formalism this means the operator must commute with the action of the algebra, while in the anyon formalism we can invoke Schur's lemma which will constrain the action on the outgoing anyon.\footnote{This can be generalised to any quasi-triangular Hopf $H$ with a corresponding anyonic theory. In the spin formalism the operator must commute with the action of $H$, which by Schur's lemma will constrain the operator's action on the outgoing anyon in the anyon formalism.}

Suppose we have an operator, $\mathcal{O}$, in the spin basis. It is said to have $D_{3}$ symmetry if it commutes with the action of the algebra,
\begin{equation} \label{EqnSymSpin}
[ \Pi(g), \mathcal{O} ] = 0 \hspace{1cm} \mbox{where} \hspace{1cm} \Pi = \left(\left[ \pi_{2} \ds \pi_{+} \ds \pi_{-} \right] \tp \pi_{2}^{\tp \mathcal{L}} \right) \circ \Delta^{(\mathcal{L}+1)},
\end{equation}
for all $g\in D_{3}$. This operator will have a counterpart in the anyon fusion path which we will denote $\tilde{\mathcal{O}}$. The $D_{3}$ symmetry is both sufficient and necessary. As the operator commutes with the action of the algebra Schur's lemma requires:
\begin{equation} \label{EqnSymAnyon}
\mbox{If} \quad\quad a_{\mathcal{L}} \neq a_{\mathcal{L}}' \quad\quad \mbox{then} \quad\quad \bra[a_{0}'a_{1}'...a_{\mathcal{L}}'] \tilde{\mathcal{O}} \ket[a_{0}a_{1}...a_{\mathcal{L}}] = 0.
\end{equation}
Thus for an operator, $\tilde{\mathcal{O}}$, in the fusion path basis to have a spin counterpart the last label, $a_{\mathcal{L}}$, must be invariant under the action of $\tilde{\mathcal{O}}$. Similarly, from construction the auxiliary space must be also invariant under the action of the operator and thus the first label $a_{0}$ is invariant under $\tilde{\mathcal{O}}$. The fixing of $a_{0}$ and $a_{\mathcal{L}}$ are necessary and sufficient for an operator in the fusion path basis to have a counterpart in the spin basis.\footnote{As stated previously this has a natural generalisation to other models with a Hopf algebra symmetry. Condition (\ref{EqnSymSpin}) is unchanged while Condition (\ref{EqnSymAnyon}) requires that the out-going anyon only remains of the same type. This modification is necessary when multiple copies of the same anyon can appear after the fusion of two anyons.} 
This is the same as the operator having hidden quantum group symmetry \cite{SliBai2001}.
Any such operator which acts non-trivially on $k$ neighbouring sites (in the spin basis) will have an anyon counterpart that acts on $k+1$ labels, e.g. for $h\in V_{2}\tp V_{2}$ we have the equivalence $h_{i(i+1)} \leftrightarrow \tilde{h}_{(i-1)i(i+1)}$. 
This correspondence between different operators is equivalent to Pasquier's face-vertex mapping \cite{Pasquier1988} which relies on Ocneanu cell calculus \cite{Roche1990}.
\\

\noindent
\textbf{\underline{Projection Operators and the Local Hamiltonian}} \\
\noindent
As we are dealing with a model with $D_{3}$ symmetry we expect that the global Hamiltonian will just be composed of projection operators. Additionally we only consider models with nearest-neighbour interactions, so we further restrict ourselves to projection operators on two sites. In the spin basis we have the two-site projection operator is given by,
$$ P^{(b)} = \frac{\mbox{dim}(V_{b})}{6}\sum_{g\in D_{3}} \mbox{Trace}(\pi_{b}(g^{-1})) \,\, \pi_{2}(g) \tp \pi_{2}(g). $$
By construction this local operator commutes with the action of the algebra and has a corresponding operator in the fusion path basis. We can diagrammatically determine the projection operators in the following way \cite{FTLTKWF2007},
\vspace{0.3cm}
\begin{eqnarray*}
	\tilde{P}^{(b)}_{i-1,i,i+1} \left\{\hspace{0.4cm} 
\begin{tikzpicture}[scale=1.0]
	\put (26,5){$2$}	\put (56,5){$2$}
	\put (0,-27){$a_{i-1}$}	\put (40,-27){$a_{i}$}	\put (72,-27){$a_{i+1}$}
	\draw (0.35,-0.6) -- (2.85,-0.6);
	\draw (1.00,0.1) -- (1.00,-0.6);	\draw (2.05,0.1) -- (2.05,-0.6);
\end{tikzpicture} \hspace{0.4cm} \right\}
	& = & \sum_{b'} (F^{a_{i-1}22}_{a_{i+1}})_{b'}^{a_{i}} \, \delta_{b}^{b'} \left\{\hspace{0.35cm} 
\begin{tikzpicture}[scale=1.0]
	\put (26,5){$2$}	\put (56,5){$2$}
	\put (8,-27){$a_{i-1}$}	\put (47,-15){$b'$}	\put (67,-27){$a_{i+1}$}
	\draw (0.60,-0.6) -- (2.60,-0.6);
	\draw (1.00,0.1) -- (1.525,-0.25);	\draw (2.05,0.1) -- (1.525,-0.25); 	\draw (1.525,-0.25) -- (1.525,-0.6);
\end{tikzpicture} \hspace{0.4cm} \right\} \\
	& = & \sum_{a_{i}'} \left[\left(F^{a_{i-1}22}_{a_{i+1}}\right)^{a_{i}'}_{b}\right]^{*} \left(F^{a_{i-1}22}_{a_{i+1}}\right)^{a_{i}}_{b}
\left\{ \hspace{0.4cm} 
\begin{tikzpicture}[scale=1.0]
	\put (26,5){$2$}	\put (56,5){$2$}
	\put (0,-27){$a_{i-1}$}	\put (40,-27){$a_{i}'$}	\put (72,-27){$a_{i+1}$}
	\draw (0.35,-0.6) -- (2.85,-0.6);
	\draw (1.00,0.1) -- (1.00,-0.6);	\draw (2.05,0.1) -- (2.05,-0.6);
\end{tikzpicture} \hspace{0.4cm} \right\},
\end{eqnarray*}
provided the $F$-moves are unitary. Alternatively we can write this as \cite{GATHLTW2012,VMP2011}
\begin{eqnarray*}
	\tilde{P}^{(b)}_{i-1,i,i+1}
	& = & \sum_{a_{i-1},a_{i},a_{i}',a_{i+1}} \left[\left(F^{a_{i-1}22}_{a_{i+1}}\right)^{a_{i}'}_{b}\right]^{*} \left(F^{a_{i-1}22}_{a_{i+1}}\right)^{a_{i}}_{b} \ket[..a_{i-1}a_{i}'a_{i+1}..]\bra[..a_{i-1}a_{i}a_{i+1}..].
\end{eqnarray*}
As expected this 2-site operator acts upon 3 labels in the fusion path basis and leaves the first and last anyon invariant under its action.

The original isotropic or XXX Heisenberg local Hamiltonian was defined as the exchange interaction on neighbouring sites. This was generalised to allow the strength of the interaction for spins in the $z$-direction to differ to those in the $x$-,$y$-direction resulting in the XXZ Hamiltonian below,
\begin{eqnarray}
	h
	& = & \frac{J}{2}\left(\sigma^{x} \tp \sigma^{x} + \sigma^{y} \tp \sigma^{y}\right) + \frac{J_{z}}{2}(\sigma^{z} \tp \sigma^{z}) + \left(\frac{J_{z}}{2} - J\right) I\tp I, \label{eqnXXZ} \\
	& = & \left( \begin{array}{cccc} J_{z} - J & 0 & 0  & 0 \\ 0 & -J & J & 0 \\ 0 & J & -J & 0 \\ 0 & 0 & 0 & J_{z} - J  \end{array}\right) \nonumber \\
	& = & -2J P^{(-)} + \left(J_{z} - J\right) P^{(2)}, \nonumber
\end{eqnarray}
where $\sigma^{j}$ are the usual Pauli matrices. This local Hamiltonian commutes with the action of $D_{3}$ as it is expressible in terms of projection operators. Furthermore we can use the natural anyon analogues of the projection operators to determine its equivalent operator in the fusion path basis,
\begin{eqnarray*}
	\tilde{h} & = & -2J \tilde{P}^{(-)}  + \left(J_{z} - J\right) \tilde{P}^{(2)}.
\end{eqnarray*}
This is equivalent to the known local Hamiltonian for the `spin-1' $su(2)_{4}$ model, up to a gauge transformation, mapping the anyons $(+,2,-)$ to $(0,1,2)$  \cite{GATHLTW2012,VMP2011}. The anyons 0, 1 and 2 of $su(2)_{4}$ are the analogues of the spin-0, -1 and -2 particles of $su(2)$. 

Other two-site $D_{3}$ invariant operators can be mapped between the two formalisms by,
\begin{eqnarray}
o = c_{+}P^{(+)} + c_{-}P^{(-)} + c_{2}P^{(2)}
& \Leftrightarrow & 
\tilde{o} = c_{+}\tilde{P}^{(+)} + c_{-}\tilde{P}^{(-)} + c_{2}\tilde{P}^{(2)}, \label{eqn2SiteOpCorr}
\end{eqnarray}
where $c_{k} \in \C$.

\section{Quantum chains}
To illustrate how the boundary conditions of a quantum chain affects the global symmetry we provide an account of a variety of models.
For the spin chains we use models constructible via the QISM and its variants as these are commonly associated with quasi-triangular Hopf algebras. Open spin chains with free ends are seen to be in correspondence with open anyon chains while closed models of either type are more complicated as the global symmetry can be broken. We find that among the closed models \textit{braided} models have a clear correspondence between the spin and anyonic formulations. It is then shown that while the periodic XXZ spin chain has an anyon counterpart, generic periodic spin chains do not. Likewise we show that generic periodic anyon chains have no spin chain counterparts, this includes $D_{3}$ anyons. \\

\noindent
\textbf{\underline{Open Chains}} \\
\noindent
The simplest (and somewhat trivial) example of a direct equivalence between chains is the open chain with free ends (non-interacting boundary fields) case. Whatever symmetry is contained by the local Hamiltonian is inherited by the global Hamiltonian (using the condition of coassociativity). The spin and anyon versions are of a very similar form,
\begin{equation} \label{eqnGHamOpen}
\mathcal{H} = \sum_{i=1}^{\mathcal{L} -1}h_{i(i+1)} \hspace{1cm} \Leftrightarrow \hspace{1cm} \tilde{\mathcal{H}} = \sum_{i=1}^{\mathcal{L} -1}\tilde{h}_{(i-1)i(i+1)}.
\end{equation}
These provide models with identical energy spectra. This chain does not have a quantum group symmetry \cite{KulSkl1991,PasSal1990}, however, its invariance under the action of $D_{3}$ will guarantee special degeneracies in the spin basis. To match up the degeneracies of each of the energies the spin dimension of each vector in the fusion path basis must be considered.\footnote{To assist the reader we have included the spectrum of the $\L=4$ open chain in the appendix.} Here we can see that the open XXZ chain is equivalent to the open $D_{3}$ chain or the `spin-1' $su(2)_{4}$ chain restricted to the integer sector. 

The introduction of non-trivial boundary fields will break the $D_{3}$ in either basis removing the correspondence between the two formalisms. \\

\noindent
\textbf{\underline{Braided Chains}} \\
\noindent
Closed boundary conditions are more complicated due to the interaction between the first and last sites. One type of closed model which can be realised equivalently in both the spin and anyon bases are braided models \cite{GPPR1994,KarZap1994,LinFoe1997}. These are guaranteed to have the full symmetry of the underlying algebra. In the case of the $D_{3}$ chain, a braided model in the spin formalism requires the existence of an invertible operator $b \in V_{2} \tp V_{2}$ satisfying:\footnote{These conditions have been adapted from \cite{GPPR1994,KarZap1994,LinFoe1997} to construct a model with $D_{3}$ symmetry, that is also invariant under the action of the global braiding operator, $\mathcal{B}$, but not necessarily integrable. It should be noted that we also require $h$ and $b$ to commute but for the $D_{3}$ case this is ensured by conditions 1.}
\begin{enumerate}
	\item It is invertible and expressible in terms of projection operators of $\pi_{2} \tp \pi_{2}$, i.e. commutes with the action of the algebra on the 2-fold tensor product space,
	\item It satisfies the braid equation, $b_{12}b_{23}b_{12} = b_{23}b_{12}b_{23}$,
	\item It braids the local Hamiltonian, $h_{12}b_{23}b_{12} = b_{23}b_{12}h_{23}$ and $b_{12}b_{23}h_{12} = h_{23}b_{12}b_{23}$.
\end{enumerate}
Once such an operator is found we can define a global braiding operator and global Hamiltonian,
$$ \mathcal{B} = b_{12}b_{23}....b_{(\mathcal{L}-1)\mathcal{L}} \hspace{1cm} \mbox{and} \hspace{1cm} \mathcal{H} =  \mathcal{B}^{-1} h_{12} \mathcal{B} + \sum_{i=1}^{\mathcal{L} -1}h_{i(i+1)}. $$
It follows that the global braiding operator and global Hamiltonian must commute with the action of the algebra.\footnote{As was the case with the open chain the proof of this relies on each local operator, $b_{i(i+1)}$ and $h_{i(i+1)}$, commuting with the Hopf algebra due to its coassociativity. This is discussed in \cite{GPPR1994,KarZap1994,LinFoe1997}.} The global braiding operator plays the role of a generalised translation operator, satisfying,
$$ \mathcal{B}h_{i(i+1)}\mathcal{B}^{-1} = h_{(i+1)(i+2)} \hspace{1cm} \mbox{and} \hspace{1cm} [\mathcal{B},\mathcal{H}]= 0, $$
for $1 \leq i \leq \mathcal{L} - 2$. The additional term in this model, although it acts globally, is viewed as a local interaction as it commutes with all local Hamiltonians not acting on either site $1$ or $\mathcal{L}$. Thus compared to the open chain the additional term only gives a finite correction to the energy. 
As the global Hamiltonian commutes with the action of the algebra this model has a natural anyonic counterpart. The anyonic counterpart is obtained by interchanging the local spin and anyon operators i.e. using relation (\ref{eqn2SiteOpCorr}) to obtain $h\leftrightarrow\tilde{h}$ and $b\leftrightarrow\tilde{b}$ yielding,
$$ \tilde{\mathcal{B}} = \tilde{b}_{012}\tilde{b}_{123}....\tilde{b}_{(\mathcal{L}-2)(\mathcal{L}-1)\mathcal{L}} \hspace{1cm} \mbox{and} \hspace{1cm} \tilde{\mathcal{H}} = \tilde{\mathcal{B}}^{-1} \tilde{h}_{012} \tilde{\mathcal{B}} + \sum_{i=1}^{\mathcal{L} -1}\tilde{h}_{(i-1)i(i+1)}. $$
For the XXZ chain we find many different operators satisfying conditions 2 and 3, however, only one also satisfies condition 1. This operator corresponds to the representation of the universal $R$-matrix of $D_{3}$ and gives rise to the periodic spin chain, which we discuss in the next section. The other operators, satisfying conditions 2 and 3 but not condition 1, may correspond to different anyonic theories. \\

\noindent
\textbf{\underline{The Periodic XXZ Spin Chain}} \\
\noindent
The periodic XXZ chain can be realised in the fusion path basis as it is also a braided model. This occurs because the permutation operator is also expressible in terms of projection operators and is consequently a suitable braiding operator, explicitly this is,
$$ \Pi = P^{(+)} - P^{(-)} + P^{(2)} \hspace{1cm} \mbox{where} \hspace{1cm} \Pi(v\tp w) = w \tp v. $$
This allows the use of the braided model formalism to consider periodic XXZ spin chain in the fusion path basis.

We remark that it is in general not possible to represent periodic spin chains in the fusion path basis as periodicity can break the underlying symmetry. The breaking of this underlying symmetry is related to the (lack of) cocomutativity of the quasi-triangular Hopf algebra in question. However, irrespective of whether the symmetry is broken certain bulk properties including energy per site and the central charge are consistent with the open chain \cite{Affleck1986,BCN1986}. \\

\noindent
\textbf{\underline{The Periodic $D_{3}$ Anyon Chain}} \\
\noindent
Now we consider the periodic $D_{3}$ anyon chain starting from the view point of an open chain. Using the $\mathcal{L}$ sites with the additional auxiliary space, we have the global Hamiltonian given by Equation $(\ref{eqnGHamOpen})$. We then impose periodicity in the basis by requiring that the incoming anyon is equal to the out-going anyon, i.e. $a_{0}=a_{\mathcal{L}}$, however the model itself is not yet translationally invariant. We are only considering an invariant subspace of the full Hilbert space and now have that the auxiliary space is coupled to the rest. We can calculate both the anyon and spin dimensions
\begin{eqnarray*}
	\mbox{Anyon dimension} & = & \sum_{a} N^{(aa)}_{\mathcal{L}} = 2^{\mathcal{L}} + (-1)^{\mathcal{L}}, \\
	\mbox{Spin dimension} & = & \sum_{a} N^{(aa)}_{\mathcal{L}} \mbox{dim}(V_{a}) = \frac{1}{3}\left[5\cdot 2^{\mathcal{L}} + 4\cdot (-1)^{\mathcal{L}} \right].
\end{eqnarray*}
At this stage we have that $a_{\mathcal{L}}$ is still invariant under the action of the Hamiltonian and subsequently there still exists a corresponding model in the spin basis.

To obtain the periodic anyon models as presented in \cite{FTLTKWF2007}, which are translationally invariant, we need to include the term $\tilde{h}_{(\mathcal{L}-1)\mathcal{L} 1}$ yielding the global Hamiltonian
\begin{eqnarray*}
	\tilde{\mathcal{H}} & = & \tilde{h}_{(\L-1)\L1} + \tilde{h}_{\L 12} + \sum_{i=2}^{\mathcal{L} -1}\tilde{h}_{(i-1)i(i+1)}.
\end{eqnarray*}
Once this term is included we no longer have that the out-going (now also incoming) anyon is unchanged by the Hamiltonian implying that the $D_{3}$ symmetry is lost and this model has no spin model counterpart.\footnote{Alternatively we could have considered a $\mathcal{L}+1$ site model and required $a_{0}=a_{\mathcal{L}}$ and $a_{1}=a_{\mathcal{L}+1}$. This would not have demonstrated as clearly how the $D_{3}$ invariance is lost.}\footnote{We remark that while from the perspective of this article the $D_{3}$ symmetry has been lost there are other notions of $D_{3}$ symmetry which can be applied e.g. when the periodic anyon chain is viewed as living on a torus then eigenstates are classified by an associated flux, labelled by a $D_{3}$ anyon, through the torus, rather than by an out-going anyon \cite{AGLTT2011,FTLTKWF2007,GATHLTW2012}.}

The periodic anyon boundary condition for this model has yet to be studied in substantial detail. It follows that the ground-state energy density must be the same as the periodic spin case. Also, the central charge will match the periodic and open spin chains. It is of interest then to compare the low-lying excitations of the XXZ chain \cite{ABB1988,Hammer1986} with their anyon counterparts. While \cite{VMP2011} has already constructed the same model there is no discussion of its correspondence to the spin-1/2 XXZ model.

\section{Discussion}
A correspondence between quantum spin and anyon chains exists when there is the underlying symmetry of a quasi-triangular Hopf algebra present. The symmetry inherited by the global Hamiltonian from the local Hamiltonian will depend upon the choice of boundary conditions. Open and braided models have a natural correspondence between the spin and anyon formalisms. On the other hand periodic models generally do not in either formalism.
In the spin language the symmetry is present if the global Hamiltonian commutes with the action of the algebra, while in the anyon language we require that the incoming and out-going anyons to be invariant under the action of the global Hamiltonian.\\

\noindent
\underline{\textbf{Acknowledgements}} \\
The author would like to thank Eddy Ardonne for much discussion and allowing the author access to some of the unfinished article \cite{GATHLTW2012} (with thanks to the other authors of the unfinished article as well). Additionally the author thanks Karen Dancer, Holger Frahm, Andr\'e Grabinski, Joost Slingerland and Robert Weston for their advice concerning this article and related topics. Lastly the author wishes to thank the reviewers who directed the author to relevant literature and previous results. 
\noindent


\begin{thebibliography}{10}
\bibitem{OEIS}
\textit{The On-Line Encyclopedia of Integer Sequences}, published
  electronically at http://oeis.org, (2011).

\bibitem{Affleck1986}
I. Affleck, \textit{Universal Term in the Free Energy at a Critical Point and
  the Conformal Anomaly}, Phys. Rev. Lett., \textbf{56}, 746--748, (1986).

\bibitem{ABB1988}
F.C. Alcaraz, M.N. Barber and M.T. Batchelor, \textit{Conformal Invariance, the
  XXZ Chain and the Operator Content of Two- Dimensional Critical Systems},
  Ann. Phys., \textbf{182}, 280--343, (1988).

\bibitem{ABBBQ1987}
F.C. Alcaraz, M.N. Barber, M.T. Batchelor, R.J. Baxter and G.R.W. Quispel,
  \textit{Surface exponents of the quantum {$XXZ$}, {A}shkin-{T}eller and
  {P}otts models}, J. Phys. A, \textbf{20}, 6397--6409, (1987).

\bibitem{AGLTT2011}
E. Ardonne, J. Gukelberger, A.W.W. Ludwig, S. Trebst and M. Troyer,
  \textit{Microscopic models of interacting Yang-Lee anyons}, New J. Phys.,
  \textbf{13}, 045006, (2011).

\bibitem{BCN1986}
H.W.J. Bl\"ote, J.L. Cardy and M.P. Nightingale, \textit{Conformal Invariance,
  the Central Charge, and Universal Finite-Size Amplitudes at Criticality},
  Phys. Rev. Lett., \textbf{56}, 742--745, (1986).

\bibitem{ChaPreBook1994}
V. Chari and A. Pressley, \textit{A guide to quantum groups}, Cambridge
  University Press, (1994).

\bibitem{CloGau1966}
J.D. Cloizeaux and M. Gaudin, \textit{Anisotropic Linear Magnetic Chain}, J.
  Math. Phys., \textbf{7}, 1384, (1966).

\bibitem{VegGia1989}
H.J. {de Vega} and H.J. Giacomini, \textit{Intertwining vectors and the
  connection between critical vertex and {SOS} models}, J. Phys. A,
  \textbf{22}, 2759--2779, (1989).

\bibitem{WilBai1998}
M. {de Wild Propitius} and F.A. Bais, \textit{Discrete gauge theories}, In
  Particles and Fields, Eds. G. Semenoff and L. Vinet, CRM Series in
  Mathematical Physics (Springer-Verlag, New York), 353--353, (1998).

\bibitem{EFGKVBook2005}
F. Essler, H. Frahm, F. G{\"o}hmann, A. Kl{\"u}mper and V.E. Korepin,
  \textit{The one-dimensional Hubbard model}, Cambridge University Press,
  (2005).

\bibitem{FTLTKWF2007}
A. Feiguin, S. Trebst, A.W.W. Ludwig, M. Troyer, A.Y. Kitaev, Z. Wang and M.H.
  Freedman, \textit{Interacting Anyons in Topological Quantum Liquids: The
  Golden Chain}, Physical Review Letters, \textbf{98}, 160409, (2007).

\bibitem{GATHLTW2012}
C. Gils, E. Ardonne, S. Trebst, D.A. Huse, A.W.W. Ludwi, M. Troyer and Z. Wang,
  \textit{Anyonic quantum spin chains: Spin-1 generalizations and topological
  stability}, (in preperation).

\bibitem{GPPR1994}
H. Grosse, S. Pallua, P. Prester and E. Raschhofer, \textit{On a quantum group
  invariant spin chain with non-local boundary conditions}, J. Phys. A: Math.
  Gen., \textbf{27}, 4761-4771, (1994).

\bibitem{Haldane1991}
F.D.M. Haldane, \textit{``Fractional statistics'' in arbitrary dimensions: A
  generalization of the Pauli principle}, Phys. Rev. Lett., \textbf{67},
  937–940, (1991).

\bibitem{Hammer1986}
C.J. Hamer, \textit{Finite-size corrections for ground states of the XXZ
  Heisenberg chain}, J. Phys. A: Math. Gen. 19, \textbf{19}, 3335, (1986).

\bibitem{KarZap1994}
M. Karowski and A. Zapletal, \textit{Quantum Group Invariant Integrable n-State
  Vertex Models with Periodic Boundary Conditions}, Nucl. Phys. B,
  \textbf{419}, 567-588, (1994).

\bibitem{Kitaev2006}
A. Kitaev, \textit{Anyons in an exactly solved model and beyond}, Ann. Phys.,
  \textbf{321}, 2--111, (2006).

\bibitem{KBIBook1993}
V.E. Korepin, N.M. Bogoliubov and A.G. Izergin, \textit{Quantum inverse
  scattering method and correlation functions}, Cambridge University Press,
  (1993).

\bibitem{KulSkl1991}
P.P. Kulish and E.K. Sklyanin, \textit{The general {$U_q[{\rm sl}(2)]$}
  invariant {$XXZ$} integrable quantum spin chain}, J. Phys. A, \textbf{24},
  L435--L439, (1991).

\bibitem{LeiMyr1977}
J.M. Leinaas and J. Myrheim, \textit{On the theory of identical particles}, Il
  Nuovo Cimento, \textbf{37B}, 1--23, (1977).

\bibitem{LinFoe1997}
J. Links and A. Foerster, \textit{On the construction of integrable closed
  chains with quantum supersymmetry}, J. Phys. A, \textbf{30}, 2483--2487,
  (1997).

\bibitem{MajidBook1995}
S. Majid, \textit{Foundations of quantum group theory}, Cambrigde University
  Press, (1995).

\bibitem{MooSei1989}
G. Moore and N. Seiberg, \textit{Classical and quantum conformal field theory},
  Comm. Math. Phys., \textbf{123}, 171--254, (1989).

\bibitem{Orbach1958}
R. Orbach, \textit{Linear Antiferromagnetic Chain with Anisotropic Coupling},
  Phys. Rev., \textbf{112}, 309--316, (1958).

\bibitem{Pasquier1988}
V. Pasquier, \textit{Etiology of {IRF} models}, Comm. Math. Phys.,
  \textbf{118}, 355--364, (1988).

\bibitem{PasSal1990}
V. Pasquier and H. Saleur, \textit{Common structures between finite systems and
  conformal field theories through quantum groups}, Nuclear Phys. B,
  \textbf{330}, 523--556, (1990).

\bibitem{Preskill2004}
J. Preskill, \textit{Lecture Notes for Physics 219: Quantum Computation},
  Lecture Notes California Institute of Technology, (2004).

\bibitem{Roche1990}
P. Roche, \textit{Ocneanu cell calculus and integrable lattice models}, Comm.
  Math. Phys., \textbf{127}, 395--424, (1990).

\bibitem{Sklyanin1988}
E.K. Sklyanin, \textit{Boundary conditions for integrable quantum systems}, J.
  Phys. A: Math. Gen., \textbf{21}, 2375-2389, (1988).

\bibitem{STF1979}
E.K. Sklyanin, L.A. Takhtadzhyan and L.D. Faddeev, \textit{Quantum inverse
  problem method. 1}, Theor. Math. Phys., \textbf{40}, 688-706, (1979).

\bibitem{SliBai2001}
J.K. Slingerland and F.A. Bais, \textit{Quantum groups and non-Abelian braiding
  in quantum Hall systems}, Nucl. Phys. B, \textbf{612}, 229--290, (2001).

\bibitem{TTWL2008}
S. Trebst, M. Troyer, Z. Wang and A.W.W. Ludwig, \textit{A short introduction
  to Fibonacci anyon models}, Prog. Theor. Phys. Suppl., \textbf{176}, 384,
  (2008).

\bibitem{VMP2011}
V. Verbus, L. Martina and A.P. Protogenov, \textit{Chain of interacting
  $SU(2)_{4}$ anyons and quantum $SU(2)_k\times \overline {SU(2)_k}$ doubles},
  Theoretical and Mathematical Physics, \textbf{167}, 843-–855, (2011).

\bibitem{Wilczek1982}
F. Wilczek, \textit{Quantum mechanics of fractional-spin particles}, Phys. Rev.
  Lett., \textbf{49}, 957--959, (1982).

\bibitem{YanYan1966a}
C.N. Yang and C.P. Yang, \textit{One-Dimensional Chain of Anisotropic Spin-Spin
  Interactions. I. Proof of Bethe's Hypothesis for Ground State in a Finite
  System}, Phys. Rev., \textbf{150}, 321–-327, (1966).

\end{thebibliography}

\appendix
\section{Basic concepts of an anyonic system}
Anyons are particles that are generalisations of bosons and fermions. The first generalisation proposed the existence of anyons in continuous two-dimensional space and was characterised by new braiding relations. Specifically, the interchange of two indistinguishable particles was allowed to result in an arbitrary phase shift, $e^{i\theta}$, of the wave function of the particles in contrast to the restricted phases $\theta=0,\pi$ associated with bosons and fermions \cite{LeiMyr1977,Wilczek1982}. 
These braiding relations were extended to also include unitary transformations on a degenerate subspace of many particle wave function generated by the permutation of particles.
Different definitions were also put forth which described anyons in any number of spatial dimensions, such as alternate exclusion principles \cite{Haldane1991}.
In this article we are concerned with algebraic descriptions of anyonic systems given by finite monoidal categories equipped with braiding rules. These are useful for describing low-dimensional anyonic lattice models \cite{Kitaev2006,TTWL2008}.

An anyonic system will consist of a set of anyon types $\{x_{i}\}_{i=1}^{K}$ which are closed under fusion of particles, 
$$ x_{i}\tp x_{j} = \bigoplus_{k=1}^{K} n_{i,j}^{k} x_{k}$$
where the $n_{i,j}^{k}$ are non-negative integers and satisfy an associativity condition. In the case of the $D_{3}$ anyonic theory the anyon types are $\{2,+,-\}$ and the fusion rules are the same as fusion rules for tensor product decomposition. Each anyon, $x_{i}$, will have an associated quantum dimension, $d_{i}\in\mathbb{R}$, which satisfy
$$ d_{i} d_{j} = \sum_{k=1}^{K} n_{i,j}^{k} d_{k} $$
The anyonic systems considered here also require braiding rules which define the interchange of anyons and are given by a mapping, 
$$ R: x_{i}\tp x_{j} \rightarrow x_{j}\tp x_{i}.$$
This mapping defines the braiding statistics of the anyons.
For further details an introduction to the subject can be found in \cite{TTWL2008} and references therein.

\section{$F$-moves and projection operators}
We have calculated the $F$-moves though explicitly decomposing the space $V_{a} \tp V_{b} \tp V_{c}$ in the two different manners mentioned previously and by then looking at the transformations between them. The $F$-moves which deal with only one-dimensional irreps:
\begin{eqnarray*}
	(F^{abc}_{a\times b\times c})^{x}_{y} & = & \delta_{x}^{a\times b} \delta_{y}^{b\times c}	
\end{eqnarray*}
where $a,b,c \in \{+,-\}$. The $F$-moves with precisely one $2$-particle present
$$ (F^{ab2}_{2})^{x}_{y} = \delta_{x}^{a\times b} \delta_{y}^{2} \hspace{1.5cm}
(F^{a2c}_{2})^{x}_{y} = \delta_{x}^{2} \delta_{y}^{2} \hspace{1.5cm}
(F^{2bc}_{2})^{x}_{y} = \delta_{x}^{2} \delta_{y}^{b\times c} $$
where $a,b,c \in \{+,-\}$. The $F$-moves with precisely two $2$-particles present and one $+$-particle
$$\begin{array}{ccc c ccc c ccc}
	(F^{+22}_{+})^{x}_{y} &=& \delta_{x}^{2} \delta_{y}^{+} & \hspace{0.5cm} &
	(F^{2+2}_{+})^{x}_{y} &=& \delta_{x}^{2} \delta_{y}^{2} & \hspace{0.5cm} &
	(F^{22+}_{+})^{x}_{y} &=& \delta_{x}^{+} \delta_{y}^{2} \\
	(F^{+22}_{-})^{x}_{y} &=& \delta_{x}^{2} \delta_{y}^{-} &&
	(F^{2+2}_{-})^{x}_{y} &=& \delta_{x}^{2} \delta_{y}^{2} &&
	(F^{22+}_{-})^{x}_{y} &=& \delta_{x}^{-} \delta_{y}^{2} \\
	(F^{+22}_{2})^{x}_{y} &=& \delta_{x}^{2} \delta_{y}^{2} &&
	(F^{2+2}_{2})^{x}_{y} &=& \delta_{x}^{2} \delta_{y}^{2} &&
	(F^{22+}_{2})^{x}_{y} &=& \delta_{x}^{2} \delta_{y}^{2} 
\end{array}$$
Here are the $F$-moves with precisely two $2$-particles present and one $-$-particle
$$\begin{array}{ccc c ccc c ccc}
	(F^{-22}_{+})^{x}_{y} &=& \delta_{x}^{2} \delta_{y}^{-} & \hspace{0.5cm} &
	(F^{2-2}_{+})^{x}_{y} &=& -\delta_{x}^{2} \delta_{y}^{2} & \hspace{0.5cm} &
	(F^{22-}_{+})^{x}_{y} &=& -\delta_{x}^{-} \delta_{y}^{2} \\
	(F^{-22}_{-})^{x}_{y} &=& \delta_{x}^{2} \delta_{y}^{+} &&
	(F^{2-2}_{-})^{x}_{y} &=& -\delta_{x}^{2} \delta_{y}^{2} &&
	(F^{22-}_{-})^{x}_{y} &=& -\delta_{x}^{+} \delta_{y}^{2} \\
	(F^{-22}_{2})^{x}_{y} &=& -\delta_{x}^{2} \delta_{y}^{2} &&
	(F^{2-2}_{2})^{x}_{y} &=& \delta_{x}^{2} \delta_{y}^{2} &&
	(F^{22-}_{2})^{x}_{y} &=& -\delta_{x}^{2} \delta_{y}^{2} 
\end{array}$$
Here are the other $F$-moves with all $2$-particles:
$$\begin{array}{ccc c ccc}
	(F^{222}_{+})^{x}_{y} &=& \delta_{x}^{2} \delta_{y}^{2} & \hspace{1.5cm} &
	(F^{222}_{-})^{x}_{y} &=& -\delta_{x}^{2} \delta_{y}^{2}
\end{array}$$
and 
\begin{eqnarray*}
	(F^{222}_{2})^{x}_{y} 
	& = & \frac{1}{2}(\delta_{x}^{+}\delta_{y}^{+} - \delta_{x}^{+}\delta_{y}^{-} + \delta_{x}^{-} \delta_{y}^{+} - \delta_{x}^{-} \delta_{y}^{-}) + \frac{1}{\sqrt{2}} (\delta_{x}^{+} \delta_{y}^{2} - \delta_{x}^{-} \delta_{y}^{2} + \delta_{x}^{2} \delta_{y}^{+} + \delta_{x}^{2} \delta_{y}^{-})
\end{eqnarray*}
The projection operators are:
\begin{eqnarray*}
	\tilde{P}^{(+)}_{(i-1)i(i+1)} & = & n_{i-1}^{+}n_{i+1}^{+} + n_{i-1}^{-}n_{i+1}^{-} + \frac{1}{4} n_{i-1}^{2} \left(\begin{array}{ccc} 1 & 1 & \sqrt{2} \\ 1 & 1 & \sqrt{2} \\ \sqrt{2} & \sqrt{2} & 2 \end{array} \right)_{i} n_{i+1}^{2}, \\
	\tilde{P}^{(-)}_{(i-1)i(i+1)} & = & n_{i-1}^{+}n_{i+1}^{-} + n_{i-1}^{-}n_{i+1}^{+} + \frac{1}{4} n_{i-1}^{2} \left(\begin{array}{ccc} 1 & 1 & -\sqrt{2} \\ 1 & 1 & -\sqrt{2} \\ -\sqrt{2} & -\sqrt{2} & 2 \end{array} \right)_{i} n_{i+1}^{2}, \\
	\tilde{P}^{(2)}_{(i-1)i(i+1)} & = & n_{i-1}^{+}n_{i+1}^{2} + n_{i-1}^{2}n_{i+1}^{+} + n_{i-1}^{-}n_{i+1}^{2} + n_{i-1}^{2}n_{i+1}^{-} + \frac{1}{2} n_{i-1}^{2} \left(\begin{array}{ccc} 1 & -1 & 0 \\ -1 & 1 & 0 \\ 0 & 0 & 0 \end{array} \right)_{i} n_{i+1}^{2}. \\
\end{eqnarray*}
We have adopted the notation that $n_{i}^{a}$ projects onto anyon $a$ at the $i$th label, i.e. $n_{i}^{a} =\ket[..a_{i}..]\bra[..a_{i}..]$, and the vector $(x,y,z)_{i}^{T}$ corresponds to $x\ket[..+_{i}..] + y\ket[..-_{i}..] + z\ket[..2_{i}..]$.

\section{Energies for the open chain with free ends}
There is an equivalence between the spin and anyon formalism for the open chains with free ends. Here we have provided a concrete example to demonstrate this. For the open XXZ chain with four sites we set the coupling parameters to $J=1$ and $J_{z}=\cosh(\frac{2i\pi}{3})$. Furthermore, we restrict our auxiliary space to $V_{+}$, implying a spin dimension of $2^{4}=16$ and anyon dimension of $\sum_{a}N_{3}^{(+a)} = 11$. Numerically we find the energies and multiplicities presented in Table \ref{tab:OpenL4Spec}.\\
\vspace{-0.5cm}
\begin{table}[h]
\caption{\label{tab:OpenL4Spec}The spectrum of the $\L=4$ open chain, restricted to the $V_{+}$ component of the auxiliary space, with $J=1$ and $J_{z}=\cosh(\frac{2i\pi}{3})$. Spin Mult. and Anyon Mult. refer respectively to the multiplicity of the energy occurring in the spin and anyon formalisms. The symmetry sector is defined as the outgoing anyon or the subspace appearing in the decomposition of $\pi_{+} \tp \pi_{2}^{\tp 4}$ which the eigenstate belongs to.}
\begin{tabularx}{\textwidth}{CCCC} \hline \hline
	Symmetry sector & Energy & Spin Mult. & Anyon Mult.  \tabularnewline \hline
	$\pi_{+}$
	& $-3.9050$ & $1$ & $1$ \tabularnewline 
	& $1.8924$ & $1$ & $1$ \tabularnewline 
	& $5.3781$ & $1$ & $1$ \tabularnewline \hline
	$\pi_{-}$
	& $-3.1196$ & $1$ & $1$ \tabularnewline 
	& $1.1218$ & $1$ & $1$ \tabularnewline
	& $5.3632$ & $1$ & $1$ \tabularnewline \hline
	$\pi_{2}$
	& $-0.0665$ & $2$ & $1$ \tabularnewline
	& $1.8290$ & $2$ & $1$ \tabularnewline
	& $5.4320$ & $2$ & $1$ \tabularnewline
	& $5.5365$ & $2$ & $1$ \tabularnewline
	& $9.3655$ & $2$ & $1$ \tabularnewline \hline \hline
\end{tabularx}
\end{table}\\

\vspace{-0.5cm}
\noindent
We see that the eigenspectra of the two formalisms are the same and that the multiplicity in the spin formulation is simply the product of the dimension of irrep of the symmetry sector and the anyon multiplicity. Projection onto the $\pi_{a}$ symmetry sector in spin picture is achieved by applying the global projection operator $\frac{\mbox{dim}(V_{a})}{6}\sum_{g\in D_{3}} \mbox{Trace}(\pi_{a}(g^{-1})) \Pi(g)$.

\end{document}